\documentclass[runningheads]{llncs}
\usepackage{graphicx}
\usepackage{mathtools}
\usepackage{amsmath}
\usepackage{booktabs}
\usepackage{hyperref}   
\usepackage{url}    
 
\usepackage[utf8]{inputenc}
\usepackage{algorithm}
\usepackage[noend]{algpseudocode}
\usepackage{wrapfig}

\usepackage{ amssymb }
\makeatletter
\newcommand{\printfnsymbol}[1]{%
  \textsuperscript{\@fnsymbol{#1}}%
}
\makeatother
\usepackage{bbding}
\hypersetup{
    colorlinks=true,
    linkcolor=blue,
    filecolor=magenta,      
    urlcolor=magenta,
}
 
\begin{document}
\title{CFEA: Collaborative Feature Ensembling Adaptation for Domain Adaptation in Unsupervised Optic Disc and Cup Segmentation}
\titlerunning{Collaborative Feature Ensembling Adaptation}

\author{Peng Liu\thanks{Equal contribution}\inst{1} \and
Bin Kong\printfnsymbol{1}\inst{2} \and
Zhongyu Li\inst{3} \and
Shaoting Zhang\inst{4} \and
Ruogu Fang\Envelope\thanks{Corresponding author: ruogu.fang@bme.ufl.edu}\inst{1}}
\authorrunning{Liu et al.}
%
\institute{
    J. Crayton Pruitt Family Dept. of Biomedical Engineering, University of Florida, Gainesville, FL, USA\\ \and
    Department of Computer Science, UNC Charlotte, Charlotte, NC, USA\\ \and
    School of Software Engineering, Xi'an Jiaotong University, Xi'an, China\\
    \and
   Sensetime Research \\ 
}

\maketitle              

\begin{abstract}
Recently, deep neural networks have demonstrated comparable and even better performance with board-certified ophthalmologists in well-annotated datasets.  However, the diversity of retinal imaging devices poses a significant challenge: domain shift, which leads to performance degradation when applying the deep learning models to new testing domains. In this paper, we propose a novel unsupervised domain adaptation framework, called Collaborative Feature Ensembling Adaptation (CFEA), to effectively overcome this challenge. Our proposed CFEA is an interactive paradigm which presents an exquisite of collaborative adaptation through both adversarial learning and ensembling weights. In particular, we simultaneously achieve domain-invariance and maintain an exponential moving average of the historical predictions, which achieves a better prediction for the unlabeled data, via ensembling weights during training. Without annotating any sample from the target domain, multiple adversarial losses in encoder and decoder layers guide the extraction of domain-invariant features to confuse the domain classifier and meanwhile benefit the ensembling of smoothing weights. Comprehensive experimental results demonstrate that our CFEA model can overcome performance degradation and outperform the state-of-the-art methods in segmenting retinal optic disc and cup from fundus images. \textit{Code is available at \url{https://github.com/cswin/AWC}}.  

\end{abstract}
\section{Introduction}
Many eye diseases can be revealed by the morphology of optic disc (OD) and optic cup (OC). For instance, glaucoma is usually characterized by the large cup to disc ratio (CDR), the ratio of the vertical diameter of the cup to the vertical diameter of the disc. Currently, determining CDR is mainly performed by pathology specialists. However, it is extremely expensive to accurately calculate CDR by human experts. Furthermore, manual delineation of these lesions also introduces subjectivity, intra-  and  inter-variability. Therefore, it is essential to automate the process of calculating CDR. OD and OC segmentation are commonly adopted to automatically calculate the CDR. Nevertheless, 
both OD and OC segmentation is challenging due to the pathological lesions on the boundaries or some regions overlapping with blood vessels.



\begin{wrapfigure}{r}{0.38\textwidth}
\centering
\includegraphics[width=0.37\textwidth]{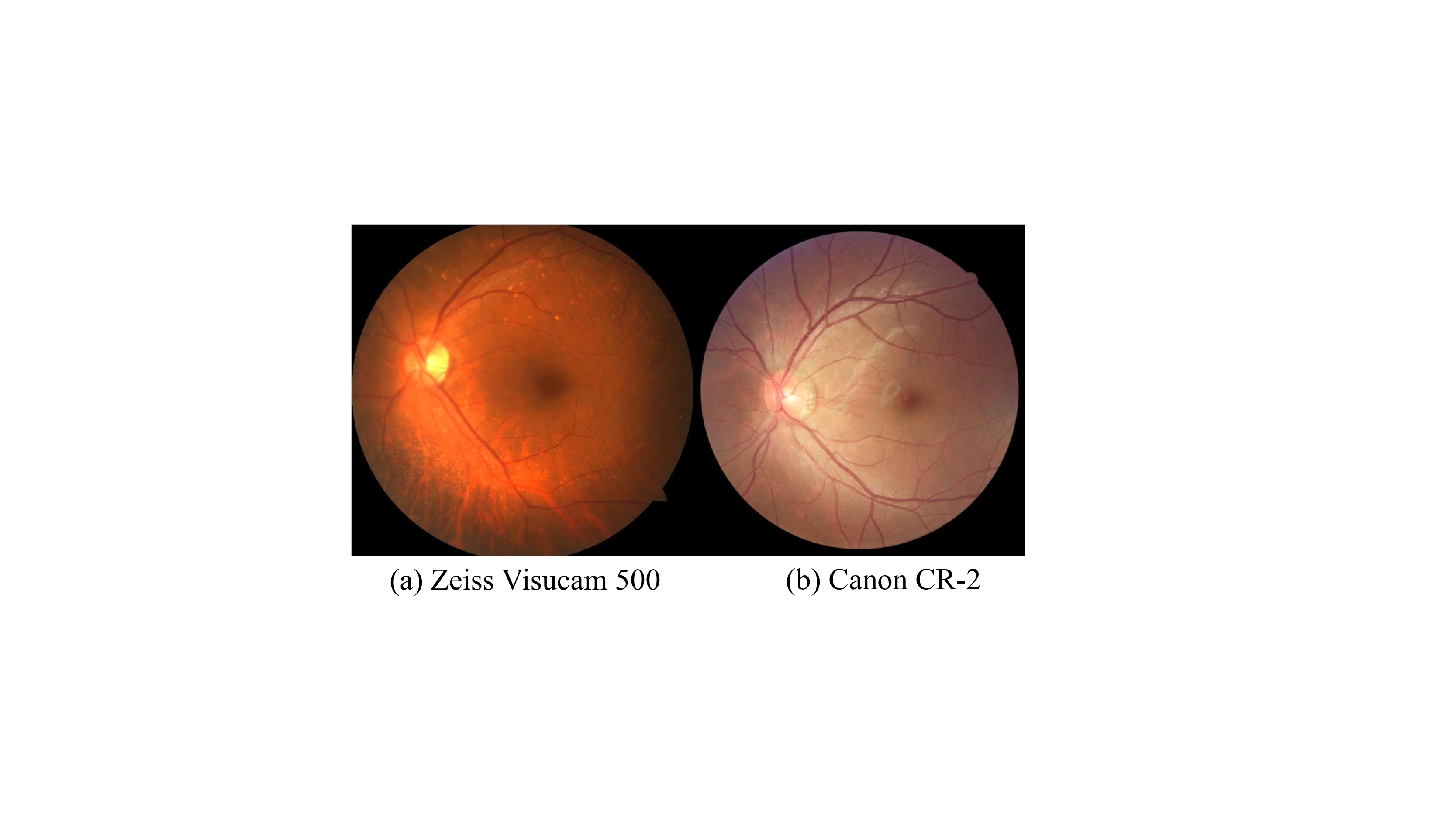}
\caption{\small Retinal fundus images collected by different fundus cameras.}
\label{fig: domains}

\end{wrapfigure}
Recently, deep learning based methods have been proposed to overcome these challenges and some of them, e.g., M-Net~\cite{fu2018joint}, have demonstrated impressive results. Although these methods tend to perform well when being applied to well-annotated datasets, the segmentation performance of a trained network may degrade severely on datasets with different distributions, particularly for the retinal fundus images captured with different imaging devices (e.g., different cameras, as illustrated in Fig.~\ref{fig: domains}). 
The variance among the diverse data domains limits deep learning's deployment in reality and impedes us from building a robust application for retinal fundus image parsing. 

To tackle this challenge, 
existing works have mainly focused on minimizing the distance between the source and target domains to align the latent feature distributions of the different domains~\cite{tzeng2017adversarial}. 
However, adversarial discriminative learning usually suffers the instability of its training.  Numerous methods have been studied to tackle this challenge. Self-ensembling~\cite{laine2016temporal} is one of them recently applied to visual domain adaptation~\cite{french2017self}. In particular, gradient descent is used to train the student network, and the exponential moving average of the weights of the student network is transferred to the teacher network after applying each training sample. The mean square difference between the outputs of the student and the teacher is used as the unsupervised loss to train the student network.  

In this paper, we propose a novel unsupervised domain adaptation framework, called Collaborative Feature Ensembling Adaptation (CFEA), to further overcome the challenges underlining in domain shift. In particular, we take the advantage of the self-ensembling, which is the time-dependent weighting to the unsupervised loss for each unlabeled sample, to stabilize the adversarial discriminative learning 
. 
Most importantly, we apply the unsupervised loss by adversarial learning not only to the output space but also to the input space or the intermediate representations of the network. 
Thus, from a complementary perspective, adversarial learning can consistently provide various model space and time-dependent weights to self-ensembling to accelerate the learning of the domain invariant features and further enhance the stabilization of adversarial learning, forming a benign collaborative circulation and unified framework. 
\begin{figure}[t]
\centering
\includegraphics[width=\textwidth]{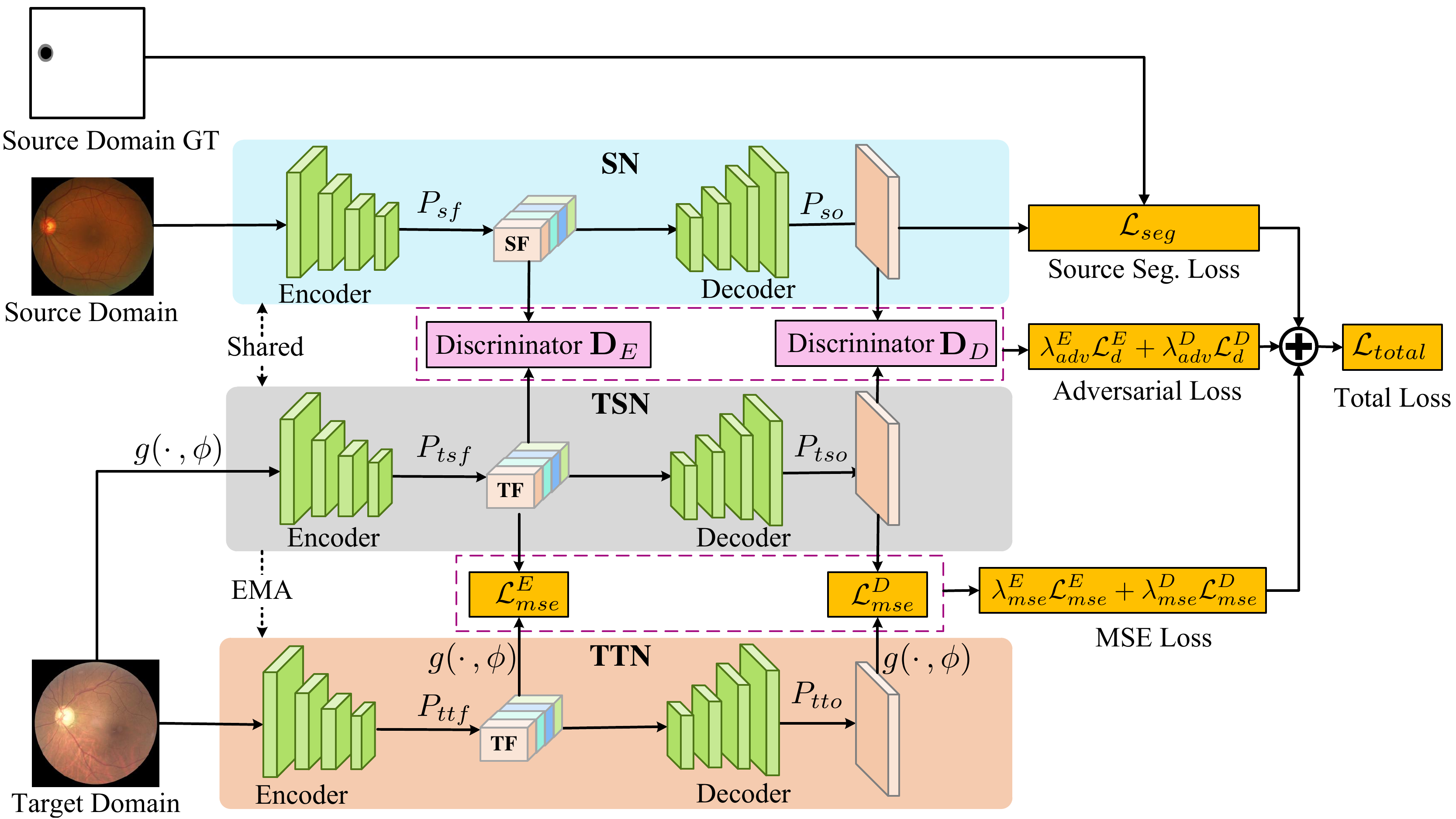}
\caption{\small Overview of the proposed model architecture.} 
\label{fig: model}
\end{figure}

The significant contributions of this paper are: (a) We propose the CFEA, a novel unsupervised domain adaptation framework, that exploits collaborative adversarial learning and self-ensembling for feature adaptation to tackle domain shift in a mutual benefit and complementary manner, thus leading to a robust and accurate model. (b) We intensify feature adaptation by applying adversarial discriminative learning in two phases of the network, i.e., intermediate representation space and output space. (c)  We evaluate the effectiveness of our CFEA on the challenging task of the unsupervised joint segmentation of retinal OD and OC. Our CFEA model can overcome performance degradation to domain shift and outperform the state-of-the-art methods.

\section{Collaborative Feature Ensembling Adaptation}
\subsection{Problem Formulation}
Unsupervised domain adaptation typically refers to the scenario: given a labeled source domain dataset with distribution $P(X_s)$ and the corresponding label $Y_s$ with distribution $P(Y_s|X_s)$, as well as a target dataset with distribution $P(X_t)$ and unknown label with distribution $P(Y_t|X_t)$, where $P(X_s)\neq P(X_t)$ , the goal is to train a model from both labeled data $X_s$ and unlabeled data $X_t$, with which the expected model distribution $P(\hat{Y}_t|X_t)$ is close to $P(Y_t|X_t)$. 

\subsection{Overview of the Proposed Method} 
As illustrated in Fig.~\ref{fig: model}, our framework mainly includes three networks, i.e., the source domain network (SN, in blue), the target domain student network (TSN, in gray) and the target domain teacher network (TTN, in orange).  Although each of the networks plays a distinctive role in guiding networks to learn domain invariant representations, all of them can interact with each other, benefit to one another, and work collaboratively as a unified framework during an end-to-end training process. SN and TSN focus on supervised learning for labeled samples from the source domain ($X_s$) and adversarial discriminative learning for unlabeled samples from the target domain ($X_t$), separately.  More importantly, we allow SN and TSN to share the weights that are sequentially learned from both labeled and unlabeled samples. The labeled samples enable the network to learn accurate segmentation predictions while the unlabeled ones bring unsupervised learning and further present a type of perturbation to regularize the model training~\cite{tarvainen2017mean}. 
Furthermore, TTN conducts the weight self-ensembling part with replicating the average weights of the TSN instead of predictions. TTN solely takes unlabeled target images as input and then the mean square difference between TSN and TTN is computed for the same target sample. Different data augmentations (e.g., adding Gaussian noise and random intensity or brightness scaling) are applied to  TSN and TTN to avoid loss vanishing issue. 

 Basically, the U-Net~\cite{ronneberger2015u} with encoder-decoder structure is employed as the backbone of each network. Since U-Net is one of the most successful segmentation frameworks in medical imaging, we expect that the results can easily generalize to other medical image analysis tasks.

\subsection{Adversarial Discriminative Learning}
\label{sec: align_gan}

We apply two discriminators at the encoder and decoder of the networks, separately, to achieve adversarial discriminative learning. Two adversarial loss functions are calculated between SN and TSN. Each of the loss calculations is performed by two steps in each training iteration: (1) train the target domain segmentation network to maximize the adversarial loss $\mathcal{L}_{adv}$, thereby fooling the domain discriminator $\textbf{D}$ to maximize the probability of the source domain feature $P_{s}$ being classified as target features:
\begin{equation}
 \begin{multlined}
\mathcal{L}_{adv}(X_s) =\mathbb{E}_{x_s\sim X_s}\log(1-\textbf{D}(P_{s})),
\end{multlined}
\label{eq: adv_general}
\end{equation}
and (2) minimize the discrininator loss $\mathcal{L}_{D}$:
\begin{equation}
 \begin{multlined}
 \mathcal{L}_{d}(X_s, X_t) =\mathbb{E}_{x_t\sim X_t}\log(\textbf{D}(P_{t}))+\mathbb{E}_{x_s\sim X_s}\log(1-\textbf{D}(P_{s})),
\end{multlined}
\label{eq: dis_general}
\end{equation}
where $P_{t}$ is the target domain feature.


\subsection{Self-ensembling}
\label{sec: self_ensemble}
In self-ensembling for domain adaptation, the training of the student model is iteratively improved by the task-specific loss,
a moving average (EMA) model (teacher) of the student model, which can be illustrated as:
\begin{equation}
 \begin{multlined}
    \Phi_t^{\prime}=\alpha\Phi_{t-1}^{\prime}+(1-\alpha)\Phi_t
\end{multlined}
\label{eq: ensembling}
\end{equation}
where $\Phi_t$ and $\Phi_t^{\prime}$ denote the paramters of the student network and the teacher network, respectively.

More specifically, at each iteration, a mini-batch of labeled source domain and unlabeled target samples are drawn from the target domain $T$. Then, the EMA predictions and the base predictions are generated by the teacher model and the student model respectively with different augmentation applied to the target samples. Afterward, a mean-squared error (MSE) loss between the EMA and target predictions is calculated. Finally, the MSE loss together with the task-specific loss on the labeled source domain data is minimized to update the parameters of the student network. Since the teacher model is an improved model at each iteration, the MSE loss helps the student model to learn from the unlabeled target domain images. Therefore, the student model and teacher model can work collaboratively to achieve robust and accurate predictions. 



\subsection{CFEA Unsupervised Domain Adaptation}
\label{sec: objective}
Unlike existing methods, our method appropriately integrates  adversarial domain confusion and self-ensembling with an encoder-decoder architecture.

\noindent\textbf{Adversarial Feature Adaptation:} Adversarial domain confusion is applied to both the encoded features and decoded predictions between source domain network (SN) and target domain student network (TSN) to reduce the distribution differences. According to Eq.~\ref{eq: adv_general} and~\ref{eq: dis_general}, this corresponds to the adversarial loss function $\mathcal{L}_{adv}^{E}$ for the encoder output of SN and TSN, and the adversarial loss function $\mathcal{L}_{adv}^{D}$ for the decoder output of SN and TSN:
\begin{align}
\label{eq: encoder_adv_framework}
\mathcal{L}_{adv}^E(X_s) =\mathbb{E}_{x_s\sim X_s}\log(1-\textbf{D}_E(P_{sf})),\\
\mathcal{L}_{adv}^D(X_s) =\mathbb{E}_{x_s\sim X_s}\log(1-\textbf{D}_D(P_{so})),
\label{eq: decoder_adv_framework}
\end{align}
where $P_{sf}\in \mathbb{R}^{W_e\times H_e\times C_e}$ and $P_{so}\in \mathbb{R}^{W_d\times H_d\times C_d}$ are the encoder and decoder outputs, respectively. $H_d$ and $W_d$ are the width and height of the decoders’ output; $ C_d $ refers to pixel categories of the segmentation result, which is three in our cases. $H_e$, $W_e$, and $ C_e $ are the width, height, channel of the encoders’ output. $\textbf{D}_E$ and $\textbf{D}_D$ are the discriminator networks for the encoder and decoder outputs, respectively.

The discriminator loss $\mathcal{L}_{d}^E$ for the encoder feature and the discriminator loss $\mathcal{L}_{d}^D$ for decoder feature are as follows:
\begin{align}
\label{eq: encoder_dis_framework}
\mathcal{L}_{d}^E(X_s, X_t) =\mathbb{E}_{x_t\sim X_t}\log(\textbf{D}_E(P_{tsf}))+\mathbb{E}_{x_s\sim X_s}\log(1-\textbf{D}_E(P_{sf})),\\
 \mathcal{L}_{d}^D(X_s, X_t) =\mathbb{E}_{x_t\sim X_t}\log(\textbf{D}_D(P_{tso}))+\mathbb{E}_{x_s\sim X_s}\log(1-\textbf{D}_D(P_{so})),
\label{eq: decoder_dis_framework}
\end{align}
where $P_{tsf}\in \mathbb{R}^{W_e\times H_e\times C_e}$ is the encoder output and $P_{tso}\in \mathbb{R}^{W_d\times H_d\times C_d}$ is the decoder output of TSN.

\noindent\textbf{Collaborative Adaptation with Self-ensembling:}
Self-ensembling is also applied to both the encoded features and decoded predictions between TSN and target domain teacher network (TTN). In this work, MSE is used for the self-ensembling. 
The MSE loss $\mathcal{L}^{E}_{mse}$ between encoder outputs of TSN and TTN, and the MSE loss $\mathcal{L}^{D}_{mse}$ between decoder outputs of TSN and TTN are as follows:
\begin{align}
\label{eq: mse_framework_enc}
\mathcal{L}^{E}_{mse}(X_t) =\mathbb{E}_{x_t\sim X_t}[\frac{1}{M}\sum_{i=1}^{M}(p^{tsf}_i-p^{ttf}_i)^2],\\
\mathcal{L}^{D}_{mse}(X_t) =\mathbb{E}_{x_t\sim X_t}[\frac{1}{N}\sum_{i=1}^{N}(p^{tso}_i-p^{tto}_i)^2].
\label{eq: mse_framework_dec}
\end{align}
where $p^{tsf}_i$, $p^{ttf}_i$, $p^{tso}_i$, and $p^{tto}_i$ denote the $i^{th}$ element of the flattened predictions ($P_{tsf}$, $P_{ttf}$, $P_{tso}$, and $P_{tto}$) of the student encoder, student decoder, teacher encoder, teacher decoder, respectively. $M$ and $N$ are the number of elements in the encoder feature and decoder output, respectively.

The same spatial-challenging augmentation $g(x,\phi)$ is used for both the teacher and student at each iteration with $g(x,\phi)$ applied to the training sample of the student and $g(x,\phi)$ applied to the predictions of the teacher, where $\phi$ is the transformation parameter. 

\noindent\textbf{Total Objective Function:} Finally, we use the dice loss as the segmentation loss for labeled images from the source domain. Combing Eq.~\ref{eq: encoder_adv_framework},~\ref{eq: decoder_adv_framework},~\ref{eq: encoder_dis_framework},~\ref{eq: decoder_dis_framework},~\ref{eq: mse_framework_enc}, and~\ref{eq: mse_framework_dec}, the total loss is obtained, which can be formulated as below. 
  
\begin{equation}
 \begin{multlined}
    \mathcal{L}_{total}(X_{s}, X_{t})=\mathcal{L}_{seg}(X_s) + \lambda^{E}_{adv}\mathcal{L}_{d}^E(X_s, X_t) + \lambda^{D}_{adv}\mathcal{L}_{d}^D(X_s, X_t) \\ +
    \lambda^{E}_{mse}\mathcal{L}^{E}_{mse}(X_t) + \lambda^{D}_{mse}\mathcal{L}^{D}_{mse}(X_t),
\end{multlined}
\label{eq: total_loss}
\end{equation}
where $\lambda^{E}_{adv}$, $\lambda^{D}_{adv}$, $\lambda^{E}_{mse}$, and $\lambda^{D}_{mse}$ balance the weights of the losses. They are cross-validated in our experiments. $\mathcal{L}_{seg}(X_s)$ is the dice segmentation loss. Based on Eq.~\ref{eq: total_loss}, we optimize the following min-max problem:
\begin{equation}
 \begin{multlined}
    \min_{f_{\phi},f_{\tilde{{\phi}}}} \max_{\textbf{D}_E,\textbf{D}_D} \mathcal{L}_{total}(X_{s}, X_{t}),
\end{multlined}
\label{eq: min_max}
\end{equation}
where $f_{\tilde{{\phi}}}$ and $f_{\phi}$ are the source domain network with trainable weight $\tilde{{\phi}}$ and target domain network with trainable weight $\phi$. 

\section{Experiments and Results}

\noindent\textbf{Data:}
Extensive experiments are conducted on the REFUGE\footnote{https://refuge.grand-challenge.org/Home/} dataset to validate the effectiveness of the proposed method. The dataset includes 400 source domain retinal fundus images (supervised training dataset) with size $2124\times2056$,  acquired by a Zeiss Visucam 500 camera, 400 labeled (testing dataset) and 400 additional unlabeled (unsupervised training dataset) target domain retinal fundus images with size $1634\times1634$ collected by a Canon CR-2 camera. As different cameras are used, the source and target domain images have totally distinct appearances (e.g., color and texture). 
The optic disc and optical cup regions were carefully delineated by the experts. All of the methods in this section are supervised by the annotations of the source domain and evaluated by the disc and cup dice indices (DI), and the cup-to-disc ratio (CDR) on the target domain.

\noindent\textbf{Data Preprocessing:} Firstly, we detect the center of optic disc by pre-trained disc-aware ensemble network~\cite{fu2018joint}, and then center and crop optic disc regions with a size of $600\times600$ for supervised training dataset and $500\times500$ for unsupervised training dataset and test dataset. This is due to the different sizes of images acquired by the two cameras. During training, all images are resized to a small size of $128\times128$ in order to adapt the network's receptive field.

\noindent\textbf{Training:} The U-Net is used for both student and teacher network. All experiments are processed on Python v2.7, and PyTorch with GEFORCE GTX TITAN GPUs.

\begin{figure}
\centering
\includegraphics[width=0.60\textwidth]{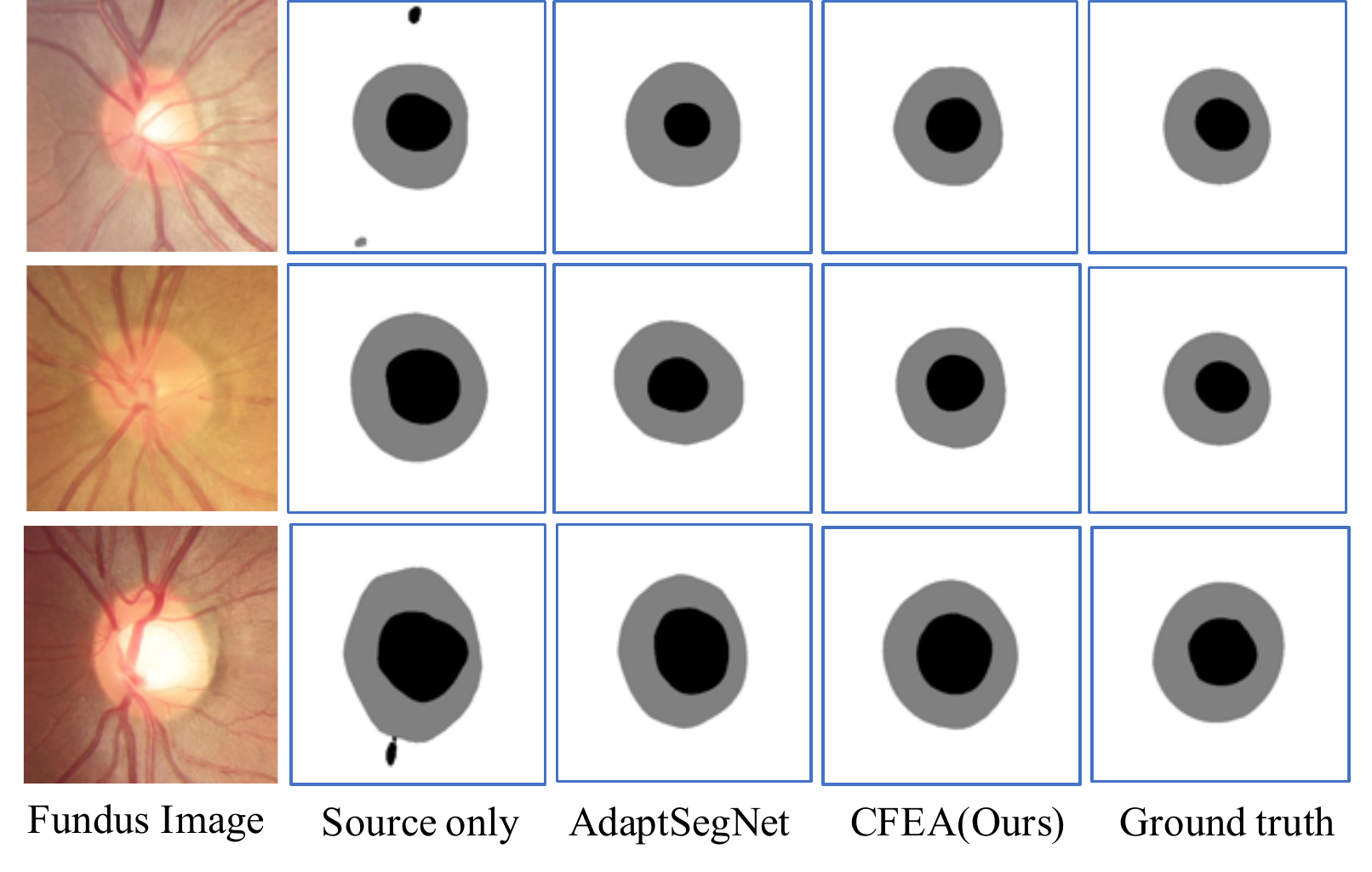}
\caption{The visual examples of optic disc and cup segmentation, where the black and gray region denote the cup and disc segmentations, respectively. From the left to right: fundus image, the model trained on source data only, baseline (AdaptSegNet~\cite{tsai2018learning}), the model trained with our domain adaptation framework, and ground truth.} 
\label{fig: mainresult}
\end{figure}
\noindent\textbf{Adaptation to different fundus cameras:}
We trained our CFEA on the source domain data acquired by Zeiss Visucam 500 camera in a supervised manner and on the target domain data acquired by Canon CR-2 camera in an unsupervised manner, simultaneously. We then evaluated our fully trained segmentation network on the test dataset, which includes 400 retinal fundus images acquired by Canon CR-2 camera. To demonstrate our method's effectiveness, we  trained the segmentation network on source domain data only in a supervised manner and then tested it on the test data. In addition, we also trained the baseline-AdaptSegNet~\cite{tsai2018learning} in the same way of training our method. AdaptSegNet~\cite{tsai2018learning} represents one of the state-of-the-art unsupervised domain adaptation methods for image segmentation, which also spplies adversarial learning for domain adaptation. The main result is shown in Table~\ref{tab: main_results}. The model trained on source data completely fails for target data. The baseline can have satisfied results on target data. By comparing our model with the baseline, as one can see, our model outperforms the state-of-the-art method consistently for OD, OC, and CDR. These results indicate the proposed framework has a capability of overcoming domain shifts, thus allowing us to build a robust and accurate model. 
\begin{table}[h]
\caption{Results of adapting source to target. We evaluate our method on 400 test images. We use three metrics to evaluate our model performance, the mean Dice coefficient for the optic cups, the mean Dice coefficient for the optic disc, and the mean absolute error for the vertical cup to disc ratio (CDR). The larger value for OD and OC means better segmentation results; for CDR, the smaller value represents better results. ``Source only'' means the model only trained on source domain in a supervised manner. AdaptSegNet~\cite{tsai2018learning} is one of the state-of-the-art unsupervised domain adaptation methods for image segmentation.}
\begin{center}
\label{tab: main_results}
\renewcommand{\arraystretch}{1}
\begin{tabular}{ c c c c c c c c}\toprule
~~~~Evaluation-Index ~~&~~~~ Source only ~~~~&~~ AdaptSegNet~\cite{tsai2018learning} ~~~~&~~CFEA(Ours)~~~~\\
\midrule

Optic Cup & 0.7317 & 0.8198 &  \textbf{0.8627} \\
\midrule
Optic Disk & 0.8532 & 0.9315 &  \textbf{0.9416 }\\
\midrule
CDR & 0.0676 & 0.0588 &  \textbf{0.0481} \\
\bottomrule
\end{tabular}
\end{center}
\end{table}

\section{Discussions and Conclusions  }

In this work, we propose a novel method CFEA for unsupervised domain adaptation of cross a diversity of retinal fundus imaging cameras. 
Our CFEA framework collaboratively combines adversarial discriminative learning and self-ensembling to obtain domain-invariant feature.
Self-ensembling can stabilize the adversarial learning and prevent the network from getting stuck in a sub-optimal solution. From a complementary perspective, adversarial learning can consistently provide various model space and time-dependent weights to self-ensembling to accelerate the learning of the domain invariant features and further enhance the stabilization of adversarial learning, forming a benign collaborative circulation and unified framework.  The collaborative mutual benefits from both adversarial feature learning and ensembling weights during an end-to-end learning process lead to a robust and accurate model.  Experimental
results demonstrate the superiority of our network over the
state-of-the-art method. Our framework needs relatively higher computational costs during the training stage to help the segmentation network to adapt to the target domain. However, in the testing stage, the computational costs will be the same as a normal U-Net network, as the images only need to go through the TTN network. Our approach is general and can be easily extended to other unsupervised domain adaptation problems. For the future work, we will conduct the extensive ablation study of the student and teacher network and the verification study of weight sharing between the SN and TSN networks.

\section{Acknowledgments}
Research reported in this publication is partially supported by the National Science Foundation under Grant No. IIS-1564892, the University of Florida Informatics Institute Junior SEED Program (00129436), the University of Florida Informatics Institute SEED Funds,  and the UF Clinical and Translational Science Institute, which is supported in part by the NIH National Center for Advancing Translational Sciences under award number UL1 TR001427. The content is solely the responsibility of the authors and does not necessarily represent the official views of the National Institutes of Health and the National Science Foundation. 
%
%
 \bibliographystyle{splncs04}
 \bibliography{refer}
\end{document}